\documentclass[10pt,a4j]{article}
\usepackage[dvips]{graphicx}
\usepackage{mathrsfs}
\usepackage{amsthm} 
\usepackage{amsmath}
\usepackage{amssymb}
\usepackage{amsfonts}
\usepackage{yhmath}
\usepackage{wrapfig}
\usepackage[dvips]{color}
\usepackage{cite}
\usepackage{array}
\usepackage{tabularx}
\usepackage[sectionbib]{chapterbib}
\usepackage{threeparttable}
\usepackage{chemarrow}
\usepackage{ascmac}
\usepackage{framed}
\definecolor{shadecolor}{gray}{0.90}
\begin{document}

\renewcommand{\figurename}{\small{Fig.}~}
\renewcommand{\labelitemi}{}
\renewcommand{\thefootnote}{$\dagger$\arabic{footnote}}
\renewcommand{\footnoterule}{% 
  \vspace{2pt}
\flushleft\rule{6.154cm}{0.4pt}
  \vspace{4pt}
}
\pagestyle{plain}

%%%%%%%%%%%%%%%%%% \UTF{0096}{\UTF{0095}¶ %%%%%%%%%%%%%%%%%
\begin{flushright}
\textit{Cyclic Bonds}
\end{flushright}
\vspace{1mm}

\begin{center}
\setlength{\baselineskip}{25pt}{\LARGE\textbf{Cyclic Bonds in Branched Polymers}}
\end{center}
\vspace{0mm}
\setlength{\baselineskip}{15pt}{\large\textbf{Reexamination of the Basic Assumptions of the Gelation Theory through Comparison with the Ilavsky-Dusek Observations}}
\vspace*{5mm}
\begin{center}
\large{Kazumi Suematsu} \vspace*{2mm}\\
\normalsize{\setlength{\baselineskip}{12pt} 
Institute of Mathematical Science\\
Ohkadai 2-31-9, Yokkaichi, Mie 512-1216, JAPAN\\
E-mail: suematsu@m3.cty-net.ne.jp,  Tel/Fax: +81 (0) 593 26 8052}\\[8mm]
\end{center}

%%%%%%%%%%%%%%%%%%
\hrule
\vspace{0mm}
\begin{flushleft}
\textbf{\large Abstract}
\end{flushleft}
\setlength{\baselineskip}{13pt}
In the theory of gelation it has been implicitly assumed that (I) a cyclic bond is a finite bond that returns to itself; (II) cyclic bonds distribute at random in network structures. In this paper these two assumptions are reexamined from a new point of view. The physical soundness of the assumptions is assessed through comparison with experimental observations.
\\[-3mm]
\begin{flushleft}
\textbf{\textbf{Key Words}}:
\normalsize{Finiteness of Cyclic Bonds/ Random Distribution of Cyclic Bonds}\\[3mm]
\end{flushleft}
\hrule
\vspace{5mm}
%%%%%%%%%%%%%%%%%% Introduction
In developing the theory of branching processes\cite{Ilavsky, Matejka, Faliagas, Spouge, Kazumi}, we have introduced two basic assumptions: (I) a cyclic bond is a finite link and (II) cyclic bonds distribute randomly in network structures. These two assumptions have served as the basis of the theory of gelation, leading us to the understanding of various phenomena that occur in the gelling processes: upward shift of gel points, existence of critical dilution, and occurrence of permanent sol. In this paper, we reinvestigate the above basic assumptions in order to lay those on firm foundation.

\section{Theoretical Background}
\subsection{On Assumption I}
In the theory of branching process, it has been implicitly assumed that a cyclic bond is a finite link, because it returns to the starting point. To date this statement has been accepted as a theorem. However, when analyzed in detail, it is seen that the statement is not necessarily self-evident. To make this issue clearer, an example of a branched structure involving one cyclic bond is illustrated in Fig. \ref{cycle} for the multiple link system of $f=3$ and $J=3$: large filled circles (\raisebox{-0.4mm}{\includegraphics[width=2.8mm]{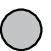}}) represent monomer units, open circles ($\circ$) functional units and the symbols (\raisebox{-0.8mm}{\includegraphics[width=3mm]{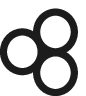}}) junction points. Suppose that a cyclization has occurred to create the bond $1-b$. Taking a look at this figure, one may raise the objection:

Why can the author affirm that a cyclic bond is finite? It is certain that the bond $1-b$ emanating from the junction point 1 returns to itself through the route $1-b-2-a-1$, but at the same time it can extend to infinity, for instance, through the route $1-b-3$, doesn't it?

That's right! But the true meaning of the cyclic bond resides in a deeper level. To discuss this problem, it is necessary to keep in mind the following two facts: (A) a cyclic bond gives no effect on the growth of cluster size; it only wastes functional units; (B) once a cyclic bond is formed, the memory of the event of cyclization is completely lost; namely, anyone of 4 bonds ($1-a$, $a-2$, $2-b$ and $b-1$) can be a cyclic bond. 
 
Because of the fact (A), there must be at least one useless, finite bond that returns to itself. Let it be the bond $1-b$ as defined above. Then disconnect the bond $1-b$, as shown in  Fig. \ref{cycle}-(B). The situation now becomes apparent. The path from the junction point 1 to 3 is still alive through another route $1-a-2-b-3$. Whether the bond $1-b$ is closed or open, the route to the junction point 3 is already constructed through the (intermolecular) bond $1-a$. It is seen that the cyclic bond is a bond not emanating from the junction point 1, but returning to itself (see Fig. \ref{cycle}-(C)). When we state that the bond 1-b is a finite bond, it is a cyclic bond, whereas when we state that the bond 1-b can lead to infinity, it is an intermolecular bond. Both the statements are true because of the fact (B). The same argument, of course, applies equally to the other bonds, $1-a$, $a-2$ and $2-b$.

\subsection{On Assumption II}
\begin{figure}[h]
\begin{center}
\vspace{4mm}
\includegraphics[width=12cm]{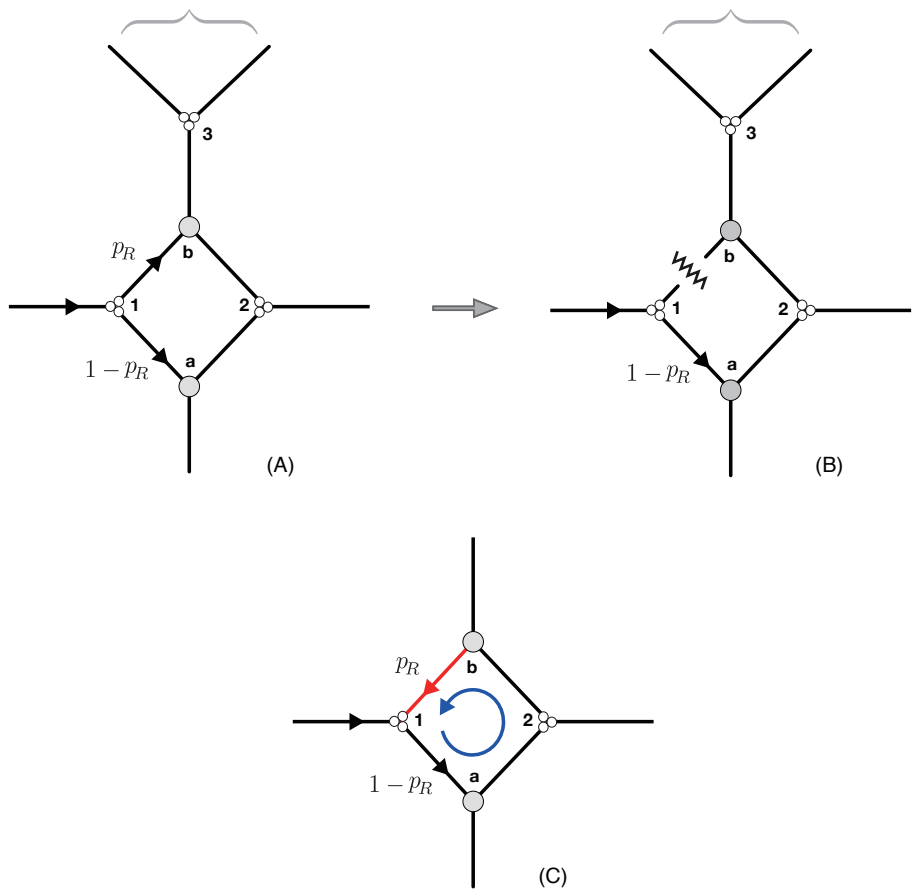}
\parbox[t]{12.7cm}{\caption{Representation of a Cyclic Bond in the Multiple Link System of $f=3$ and $J=3$.}\label{cycle}}
\end{center}
\end{figure}

As far as we consider individual clusters such as shown in Fig. \ref{cycle}, it is obvious that the validity of Assumption II is confined to the cyclic structure 1-a-2-b-1. Then it might appear that Assumption II can not be extended beyond the local structure, since individual clusters should contain different sizes, different numbers and different types of rings, so the cyclic structures change from clusters to clusters. Now the problem with which we are faced turns out to be much intricate, theoretical treatment appearing harder. However, in most theoretical treatments of branching process, what is required is the mean properties of the system; cases as to require detailed structural informations of individual clusters are rare. In such average pictures\cite{Kazumi}, every functional unit has an equal chance of cyclization or intermolecular reaction. Thus, taking the above-mentioned fact (B) into consideration, it may be concluded that there is a sufficient reason for us to accept Assumption II, the random distribution of cyclic bonds.

%%%%%%%%%%%%%%%%%% Fig-1
Our task is then to make a thorough assessment of the validity of Assumptions I and II through comparison with extensive experimental observations.

\section{Formulation of Gelling Processes}
Consider the branching process in the mixing system of the R$-$A$_{f}$ + R$-$B$_{g}$ model comprised of various monomers having two different type of functional units, $\{f_{i}M_{\text{A}_{i}} \}$ and $\{g_{j} M_{\text{B}_{j}} \}$, where $M_{\text{A}_{i}}$ and $M_{\text{B}_{j}}$ are the numbers of the A and B type monomers, respectively; $f_{i}$ and $g_{j}$ are the corresponding functionalities having positive integers with the subscripts, \textit{i} and \textit{j}, signifying $1,2,3,\cdots$. Chemical bonds can be formed only between A and B functional units. Let $[\Gamma]$ be the total molar concentration of cyclics and $V$ the system volume. Bearing in mind that each ring possesses only one cyclic bond\cite{Kazumi}, we introduce the notations:
%%%%%%%%%%%%%%%%%% eq. 1
\begin{equation}
\begin{aligned}
\bullet\hspace{3mm} &\text{fraction of cyclic bonds as against A functional units}:& & p_{R_{\text{A}}}=[\Gamma]/\left(\tfrac{1}{N_{\textit{\textsf{\hspace{-0.3mm}Av}}}V}\sum\nolimits_{i}f_{i}M_{\text{A}_{i}}\right)\equiv[\Gamma]/C_{f}\\
\bullet\hspace{3mm} &\text{fraction of cyclic bonds as against B functional units}:& & p_{R_{\text{B}}}=[\Gamma]/\left(\tfrac{1}{N_{\textit{\textsf{\hspace{-0.3mm}Av}}}V}\sum\nolimits_{j}g_{j}M_{\text{B}_{j}}\right)\equiv[\Gamma]/C_{g}\\
\bullet\hspace{3mm} &\text{fraction of A$_{i}$ functional units}:& & \chi_{\text{A}_{i}}=f_{i}M_{\text{A}_{i}}/\sum\nolimits_{i}f_{i}M_{\text{A}_{i}}\\
\bullet\hspace{3mm} &\text{fraction of B$_{j}$ functional units}:& &\chi_{\text{B}_{j}}=g_{j}M_{\text{B}_{j}}/\sum\nolimits_{j}g_{j}M_{\text{B}_{j}}\label{gel-1}
\end{aligned}
\end{equation}

\subsection{Gel Point}
Given Assumptions I and II, the number, $\alpha$, of branches leading from a given R$-$A$_f$ unit, via R$-$B$_{g}$ units, to the next R$-$A$_f$ units is simply given by
%%%%%%%%%%%%%%%%%% eq. 2
\begin{multline}
\alpha=\sum_{j}\chi_{\text{B}_{j}}\sum_{k=0}^{g_{j}-1}k\binom{g_{j}-1}{k}p_{\text{B}}^{k}(1-p_{\text{B}})^{g_{j}-1-k}\{1-P_{\text{B}}(Z|X)\}\\
\times\sum_{i}\chi_{\text{A}_{i}}\sum_{\ell=0}^{f_{i}-1}\ell\binom{f_{i}-1}{\ell}p_{\text{A}}^{\ell}(1-p_{\text{A}})^{f_{i}-1-\ell}\{1-P_{\text{A}}(Z|X)\}\label{gel-2}
\end{multline}
where $P_{\text{A}}(Z|X)=p_{R_{\text{A}}}/p_{\text{A}}$ and $P_{\text{B}}(Z|X)=p_{R_{\text{B}}}/p_{\text{B}}$ represent the conventional conditional probabilities. The critical condition occurs at $\alpha=1$, so that
%%%%%%%%%%%%%%%%%% eq. 3
\begin{equation}
(p_{\text{A}c}-p_{R_{\text{A}c}})(p_{\text{B}c}-p_{R_{\text{B}c}})=\frac{1}{(\langle f_{\text{w}}\rangle-1)(\langle g_{\text{w}}\rangle-1)}\label{gel-3}
\end{equation}
Let $\kappa=\sum\nolimits_{j}g_{j}M_{\text{B}_{j}}/\sum\nolimits_{i}f_{i}M_{\text{A}_{i}}$ be the molar ratio. Then eq. (\ref{gel-3}) may be recast in the form:
%%%%%%%%%%%%%%%%%% eq. 4
\begin{align}
p_{\text{A}c}=&\sqrt{\frac{\kappa}{(\langle f_{\text{w}}\rangle-1)(\langle g_{\text{w}}\rangle-1)}}+\frac{[\Gamma_{c}]}{C_{f}}\label{gel-4}\\
\equiv&\, p_{\text{A}c_{0}}+p_{R_{\text{A}c}}\notag
\end{align}
which is of the form: $p=p(\text{inter})+p(\text{ring})$, as expected. Note that the $p(\text{ring})$ term in eq. (\ref{gel-4}) is the quantity at $p_{c}$, so eq. (\ref{gel-4}) indicates that the gelation occurs at the point where the ratio of the intermolecular bonds to the total number of possible bonds attains the classic gel point with no rings.

To solve eq. (\ref{gel-4}) we must express $[\Gamma_{c}]$ as a function of $p_{\text{A}c}$. It is well-known that the concentration of cyclic species is held constant above the critical initial-monomer-concentration, $C_{0}\ge C_{0}^{*}$\cite{Kazumi}. For this reason we may regard the following liming solution of $C_{0}\rightarrow\infty$ as the general solution of $[\Gamma]$:
%%%%%%%%%%%%%%%%%% eq. 5
\begin{equation}
[\Gamma]=\sum_{x=1}^{\infty}\varphi_{x}\frac{[(\langle f_{\text{w}}\rangle-1)(\langle g_{\text{w}}\rangle-1)p_{\text{A}}^{2}/\kappa]^{x}}{2N_{\textit{\textsf{\hspace{-0.3mm}Av}}}\hspace{0.3mm}x} \hspace{0.7cm} (\text{for}\hspace{2mm}C_{0}\ge C_{0}^{*})\label{gel-5}
\end{equation}
Unfortunately, as emphasized previously\cite{Kazumi}, eq. (\ref{gel-5}) breaks down for $p_{\text{A}}>p_{\text{A}c_{0}}$. So we cannot link directly the limiting solution (\ref{gel-5}) with eq. (\ref{gel-4}). To resolve this problem, we expand eq. (\ref{gel-5}) about $p_{\text{A}}=p_{\text{A}c_{0}}$ to obtain
%%%%%%%%%%%%%%%%%% eq. 6
\begin{equation}
[\Gamma(p_{\text{A}})]\simeq \sum_{x=1}^{\infty}\frac{\varphi_{x}}{2N_{\textit{\textsf{\hspace{-0.3mm}Av}}}\hspace{0.3mm}x}+\sqrt{\frac{(\langle f_{\text{w}}\rangle-1)(\langle g_{\text{w}}\rangle-1)}{\kappa}}\,\,\sum_{x=1}^{\infty}\frac{\varphi_{x}}{N_{\textit{\textsf{\hspace{-0.3mm}Av}}}}(p_{\text{A}}-p_{\text{A}c_{0}})\label{gel-6}
\end{equation}
In eq. (\ref{gel-6}), we have made use of the classic relation: $(\langle f_{\text{w}}\rangle-1)(\langle g_{\text{w}}\rangle-1)p_{\text{A}c_{0}}^{2}/\kappa=1$. Let $s=(\langle f_{\text{w}}\rangle-1)(\langle g_{\text{w}}\rangle-1)/\kappa$. Then replace $p_{\text{A}}$ in eq. (\ref{gel-6}) with $p_{\text{A}c}$ and substitute the resulting equation into eq. (\ref{gel-4}) to yield
%%%%%%%%%%%%%%%%%% eq. 7
\begin{equation}
p_{\text{A}c}=\sqrt{\frac{1}{s}}\left\{\frac{1-\displaystyle\frac{\sqrt{s}}{N_{\textit{\textsf{\hspace{-0.3mm}Av}}}}\sum\nolimits_{x=1}^\infty\varphi_{x}\left(1-\tfrac{1}{2x}\right)\,\gamma_{f}}{1-\displaystyle\frac{\sqrt{s}}{N_{\textit{\textsf{\hspace{-0.3mm}Av}}}}\sum\nolimits_{x=1}^\infty\varphi_{x}\,\gamma_{f}}\right\}\label{gel-7}
\end{equation}
where $\gamma_{f}=1/C_{f}$ is the reciprocal of the initial A functional unit concentration. If we define the total functional unit concentration $C_{fg}=\tfrac{1}{N_{\textit{\textsf{\hspace{-0.3mm}Av}}}V}\big(\sum_{i}f_{i}M_{\text{A}_{i}} + \sum_{j}g_{j}M_{\text{B}_{j}}\big)$ and the total monomer concentration $C=\tfrac{1}{N_{\textit{\textsf{\hspace{-0.3mm}Av}}}V}\big(\sum\nolimits_{i}M_{\text{A}_{i}}+\sum\nolimits_{j}M_{\text{B}_{j}}\big)$, the following relationship holds between $\gamma_{f}$, $\gamma_{fg}=1/C_{fg}$ and $\gamma=1/C$:

%%%%%%%%%%%%%%%%%% eq. 8
\begin{equation}
\gamma_{f}=(1+\kappa)\gamma_{fg}=\frac{\langle f_{n}\rangle\kappa+\langle g_{n}\rangle}{\langle f_{n}\rangle\langle g_{n}\rangle}\gamma \hspace{0.7cm} (l/mol)\label{gel-8}
\end{equation}
where the subscript $n$ denotes the number average quantity\index{Number average}. 

The condition of the critical dilution can be obtained by imposing the boundary condition, $p_{\text{A}c_{0}}\le p_{\text{A}c}\le 1$, on eq. (\ref{gel-7}):
%%%%%%%%%%%%%%%%%% eq. 9
\begin{equation}
0\le\gamma_{f}\le \gamma_{f_{c}}=\frac{1-1/\sqrt{s}}{\frac{1}{N_{\textit{\textsf{\hspace{-0.3mm}Av}}}}\displaystyle\sum\nolimits_{x}\left(-1+\sqrt{s}+1/2x\right)\varphi_{x}}\label{gel-9}
\end{equation}
It is important to notice that eq. (\ref{gel-9}) is meaningful only if $\kappa\ge 1$, since otherwise the boundary condition, $p_{\text{A}c}=1$, cannot be fulfilled.

\subsection{Post-gelation}
Let $Q_{\text{A}}$ be the probability that a chosen branch emanating from the R$-$A$_{f}$ monomer unit is finite and $Q_{\text{B}}$ be the corresponding probability for the R$-$B$_{g}$ monomer unit. Given Assumptions I and II, one can readily describe a set of the recurrence relations:
%%%%%%%%%%%%%%%%%% eq. 10
\begin{equation}
\begin{split}
Q_{\text{A}}=&1-p_{\text{A}}+p_{\text{A}}\,\sum\nolimits_{j}\chi_{\text{B}_{j}}\left[P_{\text{B}}(Z|X)+(1-P_{\text{B}}(Z|X))Q_{\text{B}}^{\,g_{j}-1}\right]\\
Q_{\text{B}}=&1-p_{\text{B}}+p_{\text{B}}\,\sum\nolimits_{i}\chi_{\text{A}_{i}}\left[P_{\text{A}}(Z|X)+(1-P_{\text{A}}(Z|X))Q_{\text{A}}^{\,f_{i}-1}\right]\end{split}\label{pgel-10}
\end{equation}
The solution other than $Q_{\text{A}}=1$ or $Q_{\text{B}}=1$ is
%%%%%%%%%%%%%%%%%% eq. 11
\begin{equation}
p_{\text{A}}p_{\text{B}}(1-P_{\text{A}}(Z|X))(1-P_{\text{B}}(Z|X))\left(\sum\nolimits_{i}\chi_{\text{A}_{i}}\sum_{\ell=0}^{f_{i}-2} Q_{\text{A}}^{\,\ell}\right) \left(\sum\nolimits_{j}\chi_{\text{B}_{j}}\sum_{m=0}^{g_{j}-2} Q_{\text{B}}^{\,m}\right)=1\label{pgel-11}
\end{equation}
With $P_{\text{A}}(Z|X)=p_{R_{\text{A}}}/p_{\text{A}}$ and $P_{\text{B}}(Z|X)=p_{R_{\text{B}}}/p_{\text{B}}$, and hence $p_{R_{\text{A}}}/p_{\text{A}}\equiv p_{R_{\text{B}}}/p_{\text{B}}$ in mind, substituting $Q_{\text{A}}=Q_{\text{B}}=1$ into eq. (\ref{pgel-11}), we recover the foregoing critical condition:
%%%%%%%%%%%%%%%%%% eq. 4
\begin{equation}
p_{\text{A}_{c}}=\sqrt{\frac{\kappa}{(\langle f_{\text{w}}\rangle-1)(\langle g_{\text{w}}\rangle-1)}}+p_{R_{\text{A}c}}\tag{\ref{gel-4}$'$}
\end{equation}
with $\kappa=\sum\nolimits_{j}g_{j}M_{\text{B}_{j}}/\sum\nolimits_{i}f_{i}M_{\text{A}_{i}}\equiv p_{\text{A}}/p_{\text{B}}$ as defined above.

The sol and gel fractions of this system have the forms:
%%%%%%%%%%%%%%%%%% eq. 12
\begin{equation}
\begin{split}
W_{sol}=&\,\sum\nolimits_{i}w_{\text{A}_{i}}Q_{\text{A}}^{f_{i}}+\sum\nolimits_{j}w_{\text{B}_{j}}Q_{\text{B}}^{g_{j}}\\
W_{gel}=&\,1-W_{sol}
\end{split}\label{pgel-12}
\end{equation}
where $w_{\text{A}_{i}}$ and $w_{\text{B}_{j}}$ represent the weight fraction of the $i$th A-type monomer unit and that of the $j$th B-type monomer unit, respectively:
%%%%%%%%%%%%%%%%%% eq. 13
\begin{equation}
\begin{aligned}
w_{\text{A}_{i}}=&\frac{m_{\text{A}_{i}}M_{\text{A}_{i}}}{\sum_{i}m_{\text{A}_{i}}M_{\text{A}_{i}}+\sum_{j}m_{\text{B}_{j}}M_{\text{B}_{j}}}\\
w_{\text{B}_{j}}=&\frac{m_{\text{B}_{j}}M_{\text{B}_{j}}}{\sum_{i}m_{\text{A}_{i}}M_{\text{A}_{i}}+\sum_{j}m_{\text{B}_{j}}M_{\text{B}_{j}}}
\end{aligned}\label{pgel-wf1}
\end{equation}
and satisfy $\sum_{i}w_{\text{A}_{i}}+\sum_{j}w_{\text{B}_{j}}=1$.

\section{Examination of Results}
Eqs. (4), (7), (10) and (11) are the direct consequence of Assumptions I and II. In this paper we focus our attention to the post-gelation problem to reexamine the physical soundness of our assumptions through comparison with experiments.

\begin{figure}[h]
\begin{center}
\includegraphics[width=14cm]{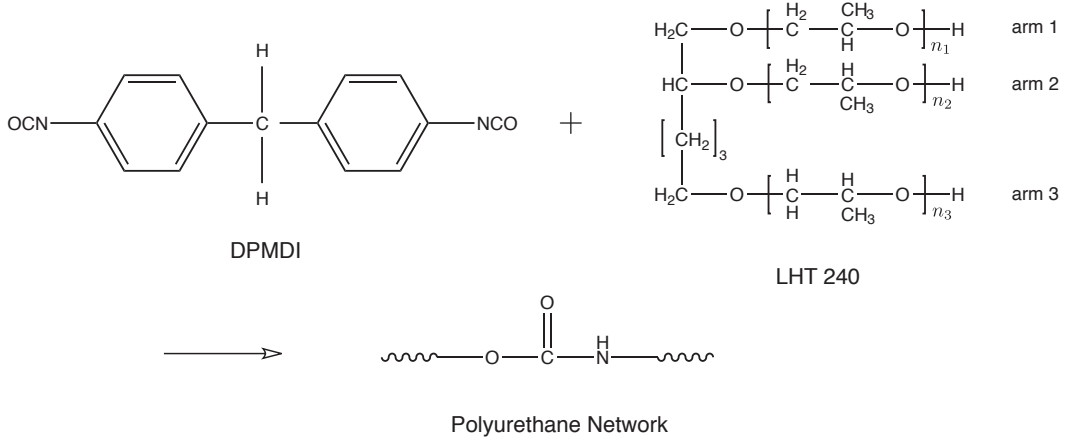}
\caption{Formation of polyurethane network by the reaction of DPMDI and LHT240.}\label{PolyUrethaneNetwork2}
\end{center}
\end{figure}

For the assessment of the above theoretical reasonings, we take up the Ilavsky-Dusek observations\cite{Ilavsky, Matejka} for the gelling process of 4,4$'$-diphenylmethane diisocyanate (DPMDI) and LHT240 at $55\sim 60\,^\circ$C in non-solvent state (see Fig. \ref{PolyUrethaneNetwork2}).

Let DPMDI be the R$-$A$_{f}$ unit ($f=2$) and LHT240 the R$-$B$_{g}$ unit, respectively, and we have $\kappa=[\text{OH}]/[\text{NCO}]$. LHT240 is prepared from 1,2,6-hexanetriol by means of the anionic polymerization and is considered to be a mixture of diols ($g_{1}=2$) and triols ($g_{2}=3$). Let $x_{1}=M_{\text{B}_{1}}/\sum_{j}M_{\text{B}_{j}}$ be the mole fraction of the diols and $x_{2}=M_{\text{B}_{2}}/\sum_{j}M_{\text{B}_{j}}$ that of the triols to the total amount of alcohols, so that $x_{1}+x_{2}=1$. According to the literature\cite{Ilavsky}, the mean functionality is $\langle g_{\text{n}}\rangle=\sum_{j}g_{j}x_{j}=2.89$. Thus $x_{1}=0.11$ and $x_{2}=0.89$. These give $\langle g_{\text{w}}\rangle=\langle g^{2}\rangle/\langle g_{\text{n}}\rangle=2.92$. Clearly $\{x_{j}\}$ can be linked with the aforementioned quantities $\{\chi_{\text{B}_{j}}\}$ by the equation: 
%%%%%%%%%%%%%%%%%% eq. 14
\begin{equation}
\chi_{\text{B}_{j}}=g_{j}x_{j}/\sum\nolimits_{j}g_{j}x_{j}\label{pgel-13}
\end{equation}
%%%%%%%%%%%%%%%%%% Fig-2
%%%%%%%%%%%%%%%%%% Fig-3
\begin{figure}[h]
\begin{center}
\includegraphics[width=11.5cm]{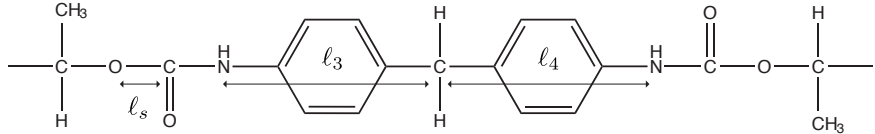}
\caption{Backbone structure of the polymer from DPMDI and LHT240.}\label{DPMDI-LHT2402D}
\end{center}
\end{figure}

\vspace*{3mm}
\noindent For the present case, $\chi_{\text{B}_{1}}=2\cdot 0.11/2.89=0.076$ and $\chi_{\text{B}_{2}}=3\cdot 0.89/2.89=0.924$. 
Solving eqs. (\ref{pgel-10})-(\ref{pgel-11}) under the condition mentioned above $(f=2,\, g_{1}=2,\, g_{2}=3)$, one has
%%%%%%%%%%%%%%%%%% eq. 15
\begin{equation}
\begin{aligned}
Q_{\text{A}}=&1+\frac{\kappa\left\{\kappa-(1+\chi_{\text{B}_{2}})(p_{\text{A}}-p_{R_{\text{A}}})^{2}\right\}}{\chi_{\text{B}_{2}}(p_{\text{A}}-p_{R_{\text{A}}})^{3}}\\
Q_{\text{B}}=&\frac{\kappa-(p_{\text{A}}-p_{R_{\text{A}}})^{2}}{\chi_{\text{B}_{2}}(p_{\text{A}}-p_{R_{\text{A}}})^{2}}
\end{aligned}\label{pgel-14}
\end{equation}
together, by eq. (\ref{gel-6}), with
%%%%%%%%%%%%%%%%%% eq. 16
\begin{align}
p_{R_{\text{A}}}=\frac{[\Gamma(p_{\text{A}})]}{C_{f}}=\left\{\sum_{x=1}^{\infty}\frac{\varphi_{x}}{2N_{\textit{\textsf{\hspace{-0.3mm}Av}}}\hspace{0.3mm}x}+\sqrt{(\langle f_{\text{w}}\rangle-1)(\langle g_{\text{w}}\rangle-1)/\kappa}\,\,\sum_{x=1}^{\infty}\frac{\varphi_{x}}{N_{\textit{\textsf{\hspace{-0.3mm}Av}}}}(p_{\text{A}}-p_{\text{A}c_{0}})\right\}\gamma_{f}\label{pgel-15}
\end{align}
Using the above quantities, the gel fraction can be calculated by the equation
%%%%%%%%%%%%%%%%%% eq. 17
\begin{equation}
W_{gel}=1-\left(w_{\text{A}}Q_{\text{A}}^{\,2}+w_{\text{B}_{1}}Q_{\text{B}}^{\,2}+w_{\text{B}_{2}}Q_{\text{B}}^{\,3}\right)\label{pgel-16}
\end{equation}
and
%%%%%%%%%%%%%%%%%% eq. 18
\begin{equation}
\begin{aligned}
w_{\text{A}}=&\frac{\langle g_{n}\rangle m_{\text{A}}}{\langle g_{n}\rangle m_{\text{A}}+f\kappa\langle m_{\text{B}n}\rangle}\\
w_{\text{B}_{j}}=&\frac{f\kappa x_{j}m_{\text{B}j}}{\langle g_{n}\rangle m_{\text{A}}+f\kappa\langle m_{\text{B}n}\rangle}
\end{aligned}\label{pgel-wf2}
\end{equation}
For the diols ($m_{\text{B}1}$) and the triols ($m_{\text{B}2}$) in question, the observed polydispersity index $\langle m_{\text{B},w}\rangle/\langle m_{\text{B},n}\rangle=1.03$\cite{Durand} gives two possible solutions: $(m_{\text{B}1}, m_{\text{B}2})=\{(359, 751), (1057, 665)\}$. In this paper we examine the solution, $(m_{\text{A}}, m_{\text{B}1}, m_{\text{B}2})=(168, 1057, 665)$.

\begin{figure}[h]
\begin{center}
\includegraphics[width=10cm]{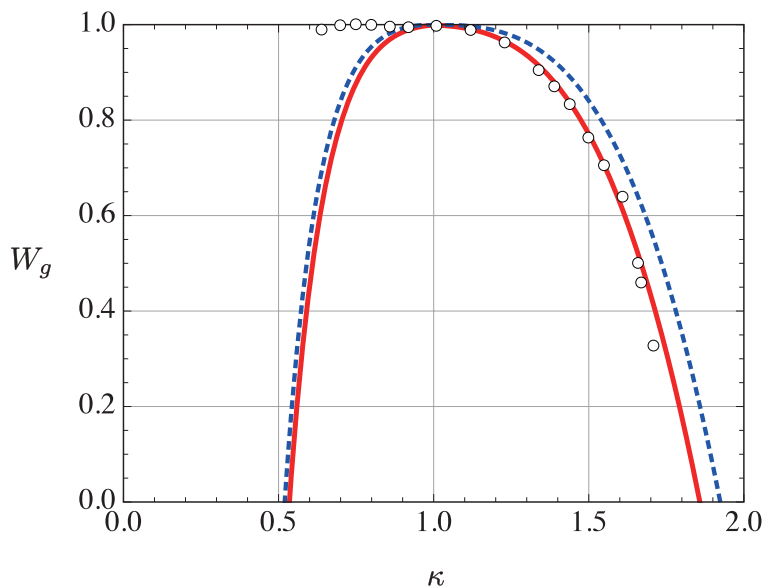}
\caption{Gel fraction as against $\kappa=[\text{OH}]/[\text{NCO}]$. Solid line ($-$): theoretical line by eqs. (\ref{pgel-14})$-$(\ref{pgel-17}); dotted line ($\cdots$): prediction by the ideal tree theory with no rings ($p_{R}=0$); ($\circ$): experimental points by Ilavsky and Dusek\cite{Ilavsky}.}\label{DPMDI-LHT240}
\end{center}
\end{figure}

Before performing our simulation, we must evaluate the relative cyclization frequency $\varphi_{x}$ defined by
%%%%%%%%%%%%%%%%%% eq. 19
\begin{equation}
\begin{split}
\varphi_x=&\left(d/2\pi^{d/2}l_{s}^{\hspace{0.3mm}d}\right)\displaystyle\int_{0}^{d/2\nu_{\hspace{-0.3mm}x}}\hspace{-2mm}t^{\frac{d}{2}-1}e^{-t}dt\\
&\nu_{x}=\langle r_{x}^{2}\rangle/l_{s}^{2}=C_{\textit{\textsf{\hspace{-0.3mm}F}}}\left(\xi_{e} x-1\right)
\end{split}\label{pgel-17}
\end{equation}
For this purpose, it is necessary to acquire the information about the effective bond number $\xi_{e}$ within the repeating unit. Set the standard bond length $l_{s}=1.36\,\text{\AA}$, and the virtual bonds, $l_{3}=l_{4}=5.69\, \text{\AA}$ (see Fig. \ref{DPMDI-LHT2402D}). Then we can calculate the required number for the $(i, j)$ arm pairs as
%%%%%%%%%%%%%%%%%% eq. 20
\begin{multline}
\xi_{i,j}=\frac{1}{1.36^{2}} \{1.36^2 + 1.36^2 + 5.69^2 + 5.69^2 + 1.36^2 + 1.36^2 + (3.3) (1.41^2 +1.53^2 +1.41^2) \\+ 1.41^2 + t\times 1.53^2
+ 1.41^2 + (3.3) (1.41^2 + 1.53^2 + 1.41^2)\}
\end{multline}
where $t=1$ for $\xi_{1,2}$, $t=4$ for $\xi_{2,3}$ and $t=5$ for $\xi_{1,3}$. And we have
%%%%%%%%%%%%%%%%%% eq. 21
\begin{equation}
\Bar{\xi}_{e}=\frac{1}{3}(\xi_{1,2}+\xi_{2,3}+\xi_{1,3})=67.92
\end{equation} 
so that $\nu_{x}=\langle r_{x}^{2}\rangle/l_{s}^{2}\simeq C_{\textit{\textsf{\hspace{-0.3mm}F}}}\,(68\,x-1)$. We set $C_{\textit{\textsf{\hspace{-0.3mm}F}}}=4.5$, the same value as that for the HMDI-LHT240 system\cite{Kazumi}. Now we can evaluate the relative cyclization frequency $\varphi_{x}$ under the condition of $d=3$. The results are summarized in Table \ref{DPMDI-LHT240-Table}. Making use of these data, we can plot the weight fraction $W_{gel}$ of gel as a function of the molar ratio $\kappa=[\text{OH}]/[\text{NCO}]$; the result is illustrated in Fig. \ref{DPMDI-LHT240} together with the experimental points by Ilavsky and Dusek\cite{Ilavsky}; the solid line represents the theoretical line based on eqs. (\ref{pgel-14})$-$(\ref{pgel-17}), and the dotted line the prediction by the ideal tree theory without rings ($p_{R}=0$). Agreement between the theory and the experiment is vey excellent for $\kappa\ge 1$, whereas marked disagreement is observed for the $\kappa\le 1$ zone. The latter phenomenon may be ascribed to side reactions, under the condition rich in $-$NCO moiety, caused by the recombination between urethane bonds $-$NH(CO)O$-$ and unreacted $-$NCO functional units to give rise to, for instance, allophanate structures\cite{Ilavsky}. The results are in good accord with the theoretical calculation based on the cascade theory by Ilavsky and Dusek\cite{Ilavsky}.

It may be concluded that the present results confirm the physical soundness of Assumptions I and II. 

%%%%%%%%%%%%%%%%%% Table 1
\begin{center}
  \begin{threeparttable}
\caption{Physicochemical parameters for DPMDI$-$LHT240.}\label{DPMDI-LHT240-Table}
\begin{tabular}{c c c}
\hline\\[-4mm]
parameters & \hspace{3mm}unit & \hspace{3mm}values \\[1mm]
\hline\\[-2mm]
molecular mass & \hspace{3mm}$m$ & \hspace{3mm} $\text{DPMDI}: m_{\text{A}}=250$ \hspace{3mm} $\text{LHT240}: \langle m_{\text{B},n}\rangle=708$\\[1mm]
specific gravity ($60\,^\circ$C) & \hspace{3mm}$\rho$ & \hspace{3mm} $\text{DPMDI}: \rho_{\text{A}}\simeq 1.0$ \hspace{3mm} $\text{LHT240}: \rho_{\text{B}}\simeq 1.0$\\[1mm]
$f$ & & \hspace{3mm} 2 \\[1mm]
$\left\langle g_{\text{n}}\right\rangle$ & & \hspace{3mm} 2.89 \\[1mm]
$\left\langle g_{\text{w}}\right\rangle$ & & \hspace{3mm} 2.92 \\[1mm]
$C_{\textit{\textsf{\hspace{-0.3mm}F}}}$ & & \hspace{3mm} 4.5\\[1mm]
$\xi_{e}$ & & \hspace{3mm} 68 \\[1mm]
$l_{s}$ & \hspace{3mm} $\left(\text{\AA}\right)$ & \hspace{3mm} 1.36 \\[2mm]
cyclization frequency & &  \\[1.5mm]
$\sum\nolimits_{x=1}^{\infty}\varphi_{x}\frac{1}{N_{\textit{\textsf{\hspace{-0.3mm}Av}}}}$ & \hspace{3mm}$(mol/l)$ & \hspace{3mm} 0.108\\[2mm]
$\sum\nolimits_{x=1}^{\infty}\varphi_{x}\frac{1}{2N_{\textit{\textsf{\hspace{-0.3mm}Av}}}\hspace{0.2mm}x}$ & \hspace{3mm}$(mol/l)$ & \hspace{3mm} 0.028\\[2mm]
\hline\\[-7mm]
   \end{tabular}
    \vspace*{2mm}
   \begin{tablenotes}
     \item 
   \end{tablenotes}
 \end{threeparttable}
\end{center}

%%%%%%%%%%%%%%%%%% Fig-4

%%%%%%%%%%%%%%%%%% References


\begin{thebibliography}{99}
\bibitem{Ilavsky}
(a) M. Ilavsky and K. Dusek. The structure and elasticity of polyurethane networks: 1. Model networks of poly(oxypropylene)triols and diisocyanate. Polymer, \textbf{24}, 981 (1983).\\
(b) M. Ilavsky and K. Dusek. Structure and elasticity of polyurethane networks. 5. Effect of diluent in the formation of model networks of poly(oxypropylene)triol and 4,4-methylenebis(phenyl isocyanate). Macromolecules, \textbf{19}, 2139 (1986).
\bibitem{Matejka}
L. Matejka and K. Dusek. Formation of polyurethane networks studied by the gel point method. Polym. Bull., \textbf{3}, 489 (1980).
\bibitem{Durand}
(a) M. Adam, M. Delsanti and D. Durand. Mechanical Measurements in the Reaction Bath during the Polycondensation Reaction, near
the Gelation Threshold. Macromolecules, \textbf{18}, 2285 (1985).\\
(b) D. Durand, F. Nveau and J. P. Busnel. Evolution of Polyurethane Gel Fraction near the Gelation Threshold. Macromolecules, \textbf{22}, 2011 (1989).
\bibitem{Faliagas} 
A. C. Faliagas. Dependence of the Cyclization Behavior of Multifunctional Network Molecules on Molecular Size. J. Polym. Sci. Part B: Polym. Phys., \textbf{43}, 861 (2005).
\bibitem{Spouge}
(a) J. L. Spouge. Equilibrium Polymer Size Distributions. Macromolecules, \textbf{16}, 121 (1983).\\
(b) J. L. Spouge. Equilibrium Ring Formation in Polymer Solutions. J. Stat. Phys., \textbf{43}, 143 (1986).
\bibitem{Kazumi}
(a) K. Suematsu. Gelation in Multiple Link System. Macromolecular Theory Simul., \textbf{12}, 476 (2003).\\
(b) K. Suematsu. Theory of Gelation: Examination of the Random Distribution Assumption of Cyclic Bonds. J. Phys. Soc. JAPAN, \textbf{75}, 064802 (2006).\\
(c) K. Suematsu. Gelation in Multiple Link System of the R$-$A$_g$ + R$-$B$_{f-g}$ Model: Examination of the Random Distribution Assumption of Cyclic Bonds. Polymer J., \textbf{38}, 1220 (2006).

\end{thebibliography}
\end{document}